\documentclass[aps,pre,floatfix,twocolumn,superscriptaddress,footinbib,showpacs,reprint]{revtex4}

\usepackage{amsmath}
\usepackage{amsfonts}
\usepackage{amssymb}
\usepackage[OT1,T1]{fontenc}
\usepackage[latin1]{inputenc}
\usepackage[english]{babel}
\usepackage[toc,page]{appendix}
\usepackage{verbatim}
\usepackage{multirow}
\usepackage{textcomp}
\usepackage{moreverb}
\usepackage{setspace}

\usepackage{epsf}
\usepackage{graphicx}
\usepackage{graphics}
\usepackage{psfrag}
\usepackage{caption}
\usepackage{subfigure}
\usepackage{placeins}
\usepackage{color}
\usepackage{subfigure}

\usepackage{mathrsfs}
\usepackage{anysize}
\usepackage{amsmath}
\usepackage{amsfonts}
\usepackage{amstext}
\usepackage{amssymb}
\usepackage{bm}
\usepackage[squaren,Gray]{SIunits}

\newcommand{\vor}{Vorono\"\i}

\begin{document}

\title{On the clustering of finite-size particles in turbulence}

\author{L.~Fiabane}
\email{lionel.fiabane@ens-lyon.org}
\affiliation{Laboratoire de Physique, ENS de Lyon, UMR CNRS 5672, Universit\'e de Lyon, France}
\author{R.~Zimmermann}
\affiliation{Laboratoire de Physique, ENS de Lyon, UMR CNRS 5672, Universit\'e de Lyon, France}
\author{R.~Volk}
\affiliation{Laboratoire de Physique, ENS de Lyon, UMR CNRS 5672, Universit\'e de Lyon, France}
\author{J.-F.~Pinton}
\affiliation{Laboratoire de Physique, ENS de Lyon, UMR CNRS 5672, Universit\'e de Lyon, France}
\author{M.~Bourgoin}
\affiliation{LEGI, UMR CNRS 5519, Universit\'e Joseph Fourier, Grenoble, France}

\begin{abstract}
We investigate experimentally the spatial distributions of heavy and neutrally buoyant particles of finite size in a fully turbulent flow. 
As their Stokes number (\emph{i.e.} ratio of the particle viscous relaxation time to a typical flow time scale) is close to unity, one may expect both classes of particles to aggregate in specific flow regions. 
This is not observed. 
Using a \vor{} analysis we show that neutrally buoyant particles sample turbulence homogeneously, whereas heavy particles do cluster. 
These results show that several dimensionless numbers are needed in the modeling (and understanding) of the behavior of particles entrained by turbulent motions.
\end{abstract}
\pacs{47.27.Gs, 47.27.T-, 82.70.-y}
\keywords{Homogeneous and isotropic turbulence, particles transport, disperse systems}

\maketitle

Turbulent flows laden with particles are widely found in industry and nature; their study is therefore of great interest and holds many fundamental aspects, issues and limits still to be explored.
One striking feature of these flows is the trend for the particles to concentrate in specific regions of the carrier flow. 
This has been observed and investigated for a long time both in experiments and simulations, and it is still widely studied (see the review paper~\cite{monchaux:2012} and references therein).
The focus is usually put on small (namely much smaller than the dissipation scale $\eta$ of the flow) and heavy particles (with a high density compared to that of the fluid), especially in numerical studies.
Because of their high specific density, the dynamics of such small and heavy inertial particles deviates from that of the carrier flow. 
Clustering phenomena are then one of the many manifestations of these inertial effects, generally attributed to the centrifugal expulsion of heavy particles  from turbulent vortices, and more recently to a sticking effect of zero-acceleration points of the carrier flow~\cite{goto:2008}. 
Other studies indicate that light particles exhibit the same trend to cluster but with different cluster geometries~\cite{calza:2008,tagawa:2012}.
Finally tracers (ought to be neutrally buoyant and much smaller than $\eta$) are used as seeds to characterize the flow dynamics and should not cluster.
As small and heavy particles, finite-size heavy particles have been found to cluster~\cite{guala:2008}.
However, the case of finite-size neutrally buoyant particles (with a diameter significantly larger than $\eta$) has never been treated to our knowledge in the context of preferential concentration.
Such particles  are known experimentally~\cite{voth:2002,qureshi:2007,brown:2009} and numerically~\cite{calza:2009,bec:2010}, to differ from tracers, with in particular different acceleration statistics.
But existing studies have focused on the dynamics of isolated particles, not on the spatial structuration of laden flows. Whether they cluster or not remains an open question.

Particles interacting with a turbulent flow are commonly characterized by their Stokes number, that is the ratio between the particle viscous relaxation time $\tau_p$ and a typical time scale of the flow.
Dealing with finite-size particles, we use the same definition as in Ref.~\cite{Schmitt:2008,Xu:2008}, using the eddy-turnover time at the scale of the particle, $\tau_d$, as the time scale of the flow, and a corrective factor $f_\phi$ based on the Reynolds number at particle scale:
$St \equiv \tau_p/\tau_d=f_\phi\phi^{4/3} (1+2\Gamma)/36$, where $\Gamma=\rho_p/\rho_f$ is the particle to fluid density ratio and $\phi=d/\eta$ is the particle diameter normalized by the dissipation scale
(note that our conclusions remain the same using a point-particle definition of the Stokes number).
This dimensionless number is often used as the key parameter to characterize particle dynamics in turbulence,
using simple Stokesian models where the dominant force acting on the particle is taken as the drag due to the difference between the particle velocity and that of the fluid.
These models predict preferential concentration of particles with non-vanishing Stokes number, with a maximal segregation for $St$ around unity~\cite{bec:2007,coleman:2009}.
This behavior is confirmed, at least qualitatively, in experiments with small and heavy particles~\cite{monchaux:2010}.
In the present study we address the case of finite-size particles (both neutrally buoyant and heavier than the fluid) and investigate the particle concentration field as a function of their Stokes number, in a homogeneous and isotropic turbulent flow.
First, we describe the experimental setup and the data processing used.
Then, we describe the results on spatial structuration for finite-size particles.
We finish with a brief discussion and conclusions.

\begin{figure}[t]
\centering
\includegraphics[width=\columnwidth]{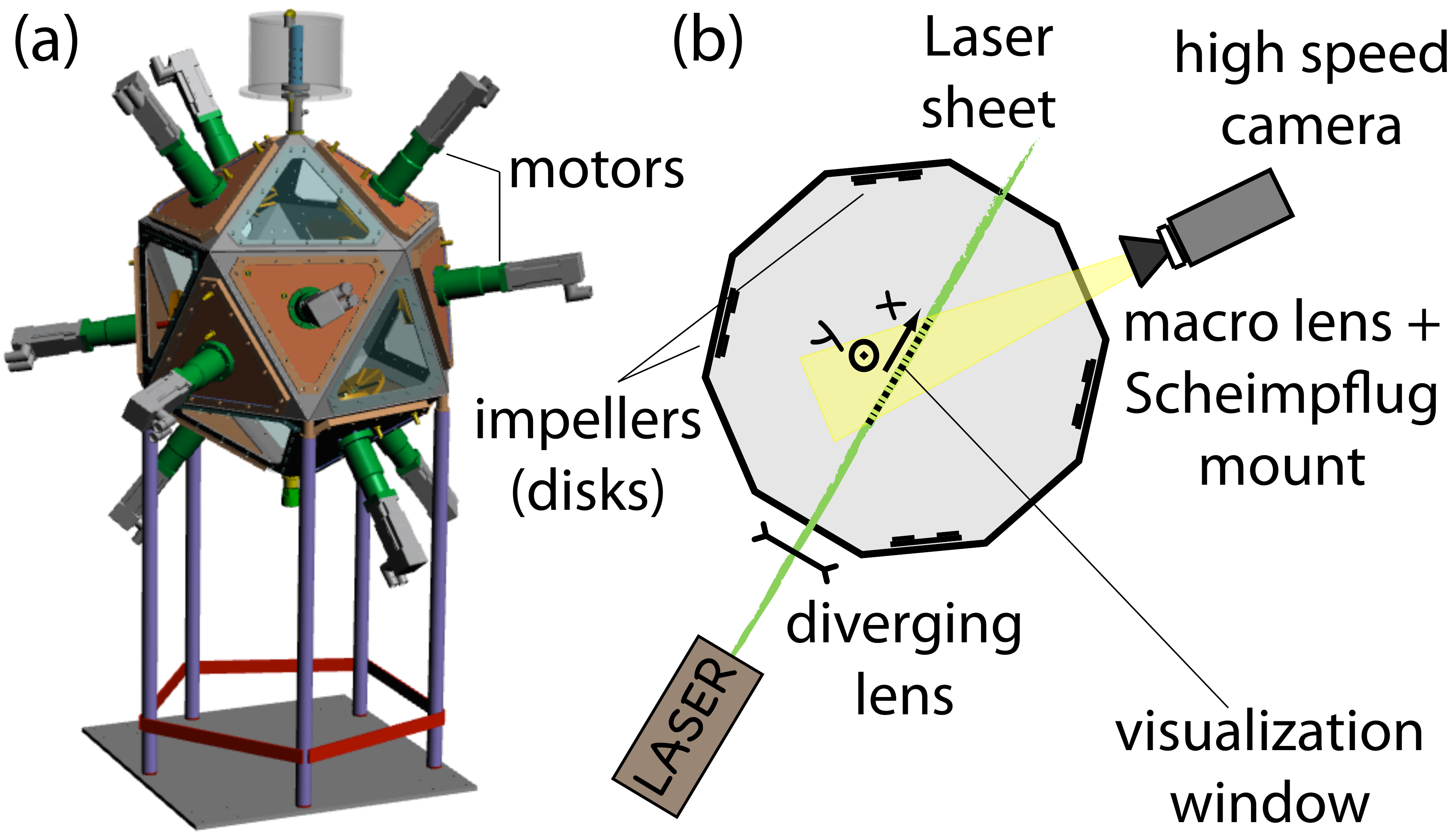}
\caption{\small{(color online) \; (a)~CAD drawing of the LEM \; (b)~Schematic upper view of the setup.}}
\label{fig:lem}
\end{figure}
\begin{table*}[t]
\small
\caption{\small{Turbulence characteristics. 
$f_\text{imp}$: rotation frequency of the 12 impellers;
$u'$: fluctuation velocity of the flow;
$\varepsilon$: energy dissipation rate; 
$\eta \equiv (\nu^3/\varepsilon)^{1/4}$ and $\tau_\eta \equiv (\nu/\varepsilon)^{1/2}$: Kolmogorov length and time scales of the flow;
$R_\lambda \equiv (15 u'^4/\nu \varepsilon)^{-1/2}$: Taylor micro-scale Reynolds number;
$St_\text{n}$ and $St_\text{h}$: Stokes numbers of neutrally buoyant  and heavy particles respectively.}}
\begin{ruledtabular}
\begin{tabular}{cccccccc}
$f_\text{imp} $ & $u' $ & $\varepsilon$ & $\eta$ & $\tau_{\eta}$ & \multirow{2}{*}{$R_{\lambda}$} & \multirow{2}{*}{$St_\text{n}$} & \multirow{2}{*}{$St_\text{h}$}\\
$\hertz$ & $\centi\meter\per\second$ & $\squaren\meter\per\second^3$ & $\micro\meter$ & $\milli\second$ &  &  & \\
\hline 
2 & 4 & 0.0016 & 159 & 24.9 & 160 & $0.38 \pm{0.02}$ & $0.25 \pm{0.04}$\\
4 & 8 & 0.0144 & 92 & 8.3 & 210 & $0.64 \pm{0.03}$ & $0.46 \pm{0.09}$\\
6 & 12 & 0.0611 & 64 & 4.0 & 260 & $0.87 \pm{0.04}$ & $0.68 \pm{0.11}$\\
8 & 17 & 0.1086 & 55 & 3.0 & 310 & $0.98 \pm{0.04}$ &$0.78 \pm{0.11}$\\
10 & 22 & 0.2087 & 47 & 2.2 & 360 & $1.11 \pm{0.05}$ & $0.92 \pm{0.11}$\\
12 & 26 & 0.3518 & 41 & 1.7 & 395 & $1.23 \pm{0.05}$ & $1.04 \pm{0.16}$
\end{tabular}
\end{ruledtabular}
\label{tab:characteristics}
\normalsize
\end{table*}
In order to study behaviors of both neutrally buoyant and heavy particles, we use water as the carrier fluid.
The turbulence is generated in the Lagrangian Exploration Module (LEM, see Fig.~\ref{fig:lem}), whose characteristics are described in detail in Ref.~\cite{zimmermann:2010} and summed up in Table~\ref{tab:characteristics}.
The LEM produces turbulence in a closed water flow forced by 12 impellers evenly distributed on the faces of an icosahedral vessel.
Here, all impellers rotate simultaneously at the same constant frequency $f_\text{imp}$ which can be varied from $2\,\hertz$ up to $12\,\hertz$; opposing impellers form counter-rotating pairs.
This produces a very homogeneous and isotropic turbulence in a significant central region of the device of the order of $(15\,\centi\meter)^3$, with the integral length of the order of $5\,\centi\meter$~\cite{zimmermann:2010}.
It also permits to obtain high Reynolds number turbulent flow with mean velocities much weaker than the fluctuations near the center of the apparatus.
Acquisitions are performed using 8 bit digital imaging at a resolution corresponding to a visualization window of the order of $16 \times 12\,\centi\meter$ in the center of the LEM. 
The visualization zone is illuminated by a $100\,\watt$ Nd:YAG pulsed laser synchronized with the camera, creating a green light sheet with millimetric thickness.
The camera is equipped with a macro lens.
A Scheimpflug mount compensates for the depth of field effects resulting from the angle between the camera and the laser sheet (see Fig.~\ref{fig:lem}).
Images are recorded at a low repetition rate of $2.5\,\hertz$ sufficient to address particle spatial distribution as we ensemble average over 2000 independent flow realizations for each experiment (particle dynamics is not addressed here).

We explore the behavior of two kinds of finite-size particles, neutrally buoyant and heavy ones.
Regarding the neutrally buoyant particles, we use expanded polystyrene particles whose density has been irreversibly adjusted (prior to actual experiments) by moderate heating so that the density ratio to water is  $1\leq \Gamma_\text{n} \leq 1.015$. 
Depending on their size, the particles can be regarded as \emph{tracers} following the carrier flow dynamics, or as \emph{finite-size particles} whose dynamics departs from that of the flow.
The limit for tracer behavior is known to be $d_\text{n}\simeq 5\eta$, whereas finite-size effects appear for $d_\text{n} > 5\eta$~\cite{qureshi:2007,brown:2009,volk:2011}.
In our case the particle diameter is $d_\text{n}=700\,\micro\meter \,\pm20\,\micro\meter$ and ranges from $4.5\eta$ to $17\eta$ (from the lowest to the highest investigated $R_{\lambda}$).
Hence, our particles transit from tracers to finite-size particles as $f_\text{imp}$ increases.
As for heavy particles, we use slightly poly-dispersed sieved glass particles with diameters $225\,\micro\meter \,\pm 25\,\micro\meter$ and a density ratio $\Gamma_\text{h}=2.5$, making them \emph{inertial particles}.
The neutrally buoyant particles cover a Stokes number from 0.38 to 1.23, while the heavy particles cover a Stokes number from 0.25 to 1.04 (see Table~\ref{tab:characteristics}).
These two ranges overlap allowing some comparison. 
As for each class of particles the diameter and density are kept constant, the Stokes number is varied by tuning the flow dissipation time scale. 
Therefore, it cannot be varied independently of the Reynolds number of the carrier flow.

\begin{figure*}[tb]
\centering
\includegraphics[width=\textwidth]{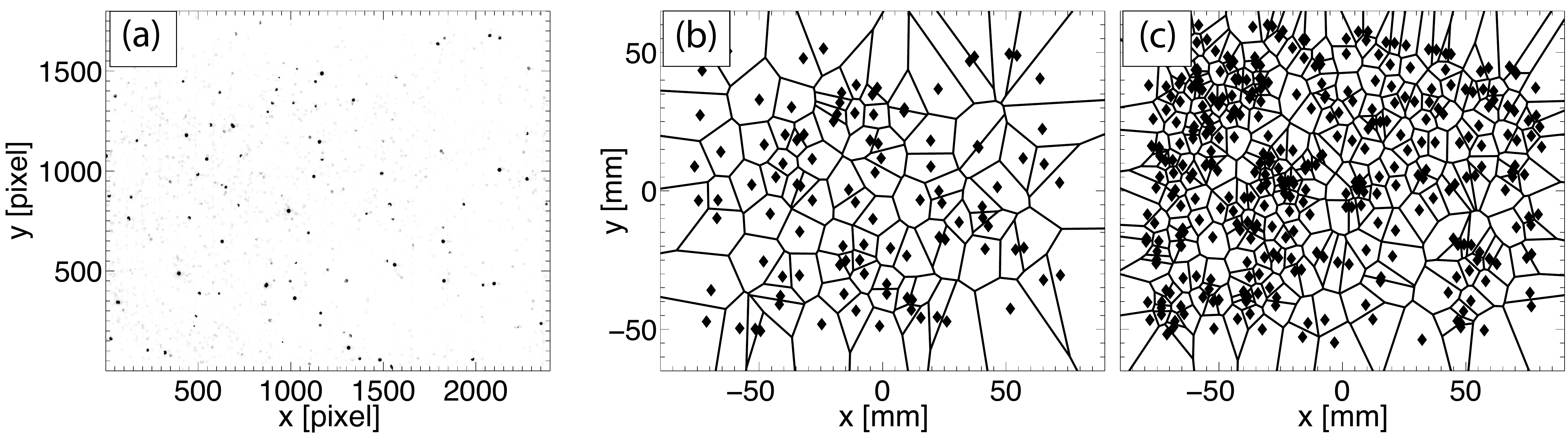}
\caption{\small{(a)~Typical raw acquired image of neutrally buoyant particles and (b)~the detected particles located in real space with the associated \vor{} diagram \; (c)~Heavy particles located in real space with the associated \vor{} diagram (corresponding raw image not shown).}}
\label{fig:tesselation}
\end{figure*}
We identify the particles on the images as local maxima with light intensity higher than a threshold, assuming in a first approximation that all the particles illuminated in the laser sheet belong to one plane.
The center of the particles is determined as the center of mass of all the pixels surrounding one local maximum.
Due to the high contrast between the light diffused by the particles and the background, changing slightly the threshold value does not significantly impact the number of detected particles,
which is of order 100 (resp. 150) for the neutrally buoyant (resp. heavy) particles 
(note that dealing with finite-size particles, the maximum authorized seeding density is drastically reduced by particles in the bulk blocking/eclipsing the image of particles in the laser sheet, compared \emph{e.g.} to experiments with small particles where thousands of particles per image are typically recorded~\cite{monchaux:2010}).
The number of detected neutrally buoyant particles remains constant in time, indicating a good stationarity of seeding concentration as expected for non-settling particles.
However large heavy particles tend to settle for low impeller rotation rates $f_\text{imp}$. The heavy particles we consider here are sufficiently large to be considered as \emph{finite size} and sufficiently small to prevent significant settling as the entrainment by the flow is still capable to keep them in suspension. Because of this limitation we did not consider bigger particles, and we did not investigate regimes where $f_\text{imp}<2\,\hertz$ (for which settling becomes important). Moreover, we make sure that the flow is already set in motion when the particles are inserted in the vessel to prevent them from settling immediately. Additionally, the number of particles per image is monitored, and experiments are repeated (after reloading particles) if too many are found to have settled (a typical experiment can run a few hours with relatively stationary seeding conditions).

The particle concentration field is investigated using \vor{} diagrams; this technique recently introduced for the investigation of preferential concentration of small water droplets in a turbulent airflow~\cite{monchaux:2010} was shown to be particularly efficient and robust to diagnose and quantify clustering phenomena. 
A given raw image, the detected particles and the associated \vor{} diagram are provided for neutrally buoyant particles in Fig.~\ref{fig:tesselation}(a,b); Fig.~\ref{fig:tesselation}(c) shows a typical \vor{} diagram for the heavy particle case.
The \vor{} diagrams give a tesselation of a two-dimensional space where each cell of the tesselation is linked to a detected particle, with all points of one cell closer to its associated particle than to any other particle.
Thus, the area of each \vor{} cell is the inverse of the local concentration of particles, \emph{i.e.} \vor{} area fields are a measure of the local concentration fields at interparticle length scale.
To compare the results of experiments made with different amount of detected particles per image, the \vor{} area is normalized using the average \vor{} area $\bar{A}$ defined as the mean particles concentration inverse, independent of the spatial organization of the particles.
Therefore, we focus in the rest of the letter on the distribution of the normalized \vor{} area $\mathcal{V}\equiv A/\bar{A}$.
Clustering properties are quantified by comparing the probability density function (PDF) of \vor{} cell areas obtained from the experiments to that of a synthetic random Poisson process (RPP) whose shape is well approximated by a Gamma distribution~\cite{ferenc:2007}.

\begin{figure}[b]
\centering
\includegraphics[width=\columnwidth]{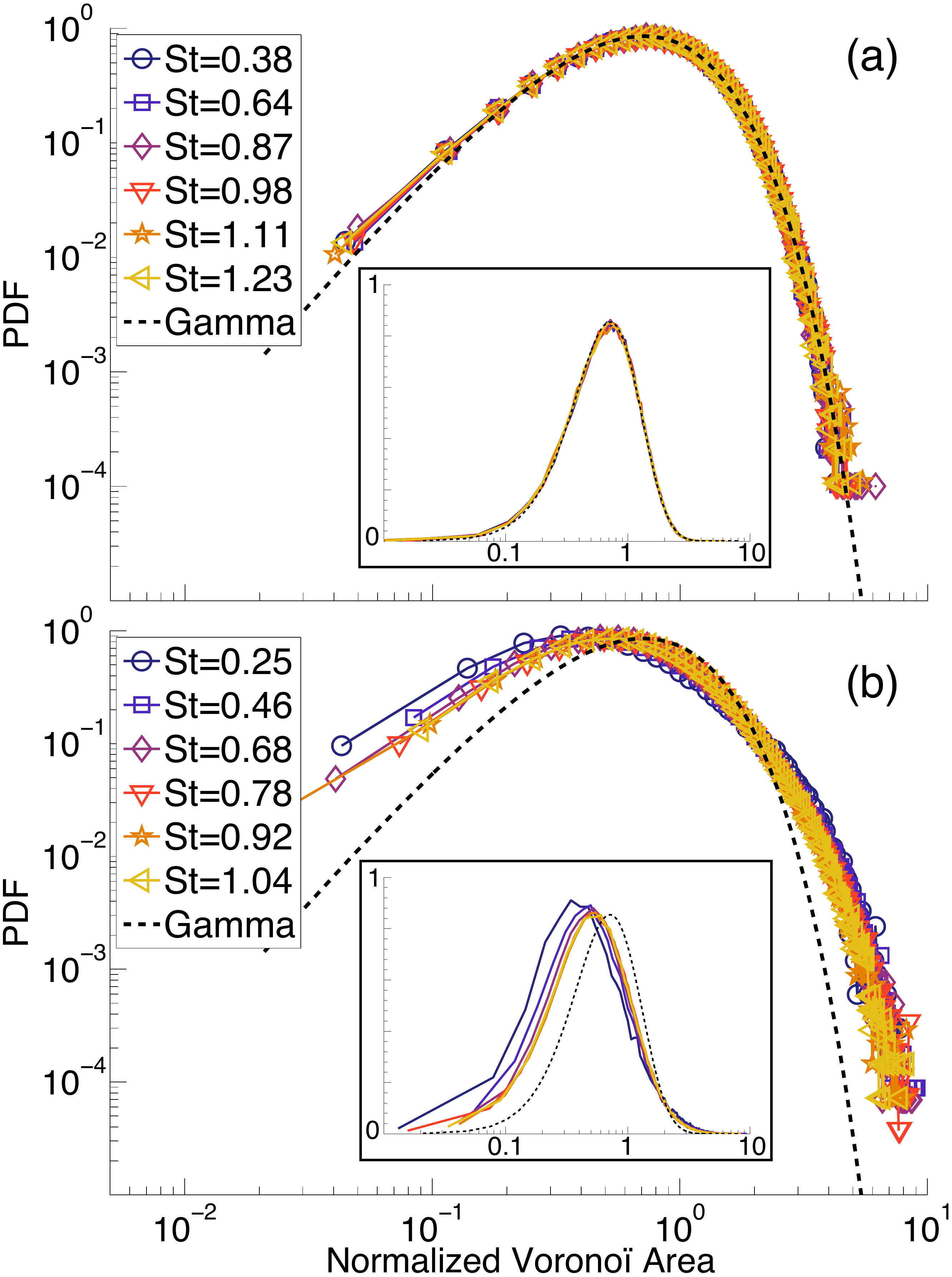}
\caption{\small{(color online) \; Superposition of the normalized \vor{} area PDFs for six experiments with varying Stokes number (plain lines) and a Gamma distribution (dashed line). Inserts represent the same results with linear ordinates. \quad (a)~Neutrally buoyant particles \; (b)~Heavy particles.}}
\label{fig:voro_pdf}
\end{figure}
The PDFs of \vor{} cell areas for the different experiments described in Table~\ref{tab:characteristics} are plotted in Fig.~\ref{fig:voro_pdf}, as well as the Gamma distribution approximation for a RPP. 
In the case of neutrally buoyant particles, all PDFs collapse and no Stokes number dependency is found. 
An important finding of the present work is that these PDFs are almost undistinguishable from the RPP distribution, meaning large neutrally buoyant particles do not exhibit any preferential concentration whatever their Stokes number.
In the case of heavy particles, the PDFs clearly depart from the RPP distribution, with higher probability of finding depleted regions (large \vor{} areas) and concentrated regions (small \vor{} areas), which is the signature of preferential concentration.
Furthermore, the shape of the PDF clearly depends on experimental parameters ($St_\text{h}$ and/or $R_\lambda$).
Interestingly, this dependence is stronger for the small \vor{} areas tails, whereas the tail for large \vor{} areas (corresponding to depleted regions) appears to be more robust. 
This was also observed for small inertial particles~\cite{monchaux:2010,obligado:2011}.
\begin{figure}[tb]
\centering
\includegraphics[width=\columnwidth]{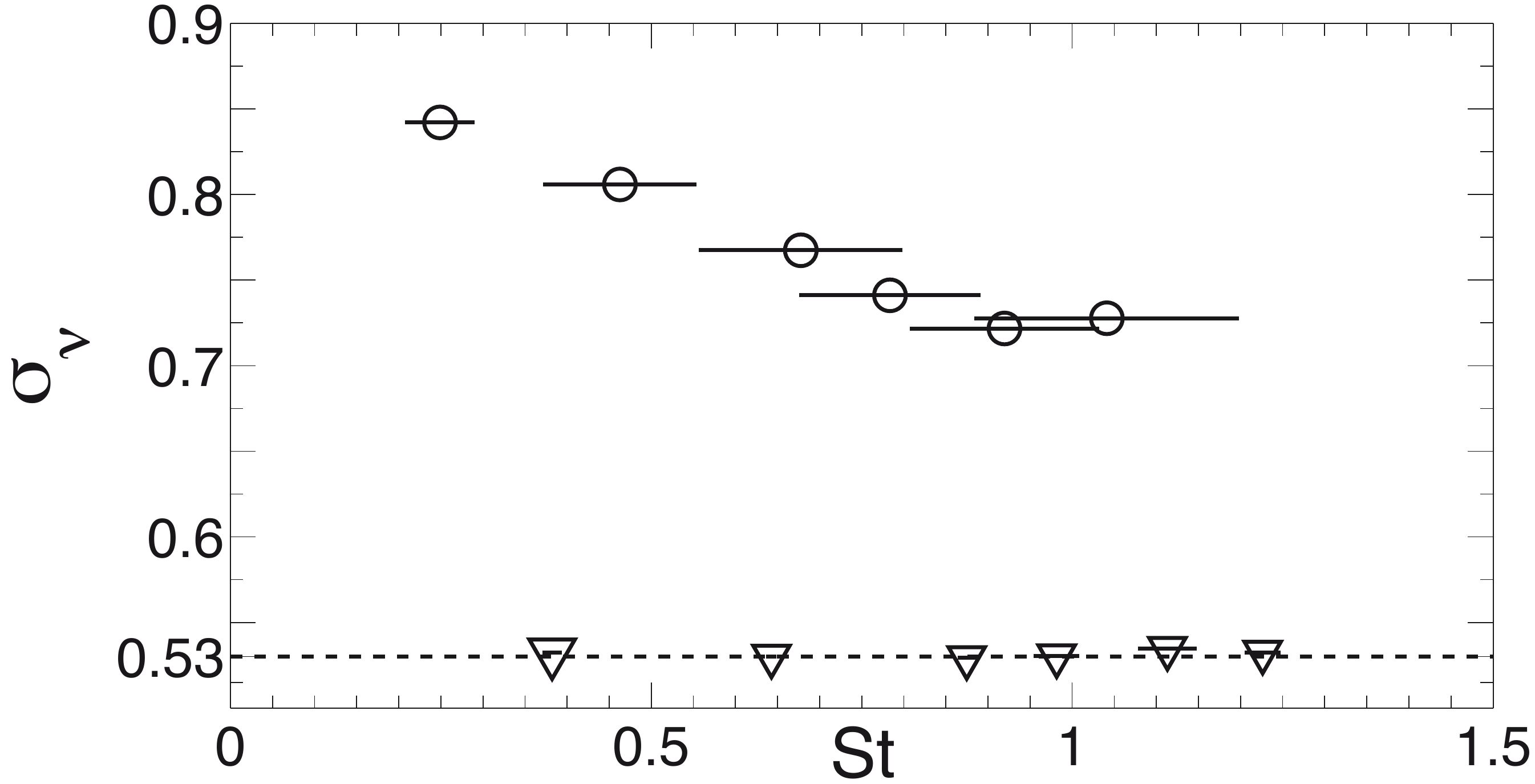}
\caption{\small{Standard deviations of the normalized \vor{} areas versus the Stokes number with error bars (plain lines) for $\triangledown$: neutrally buoyant and $\circ$: heavy particles, to compare with $\sigma_\mathcal{V}^\text{RPP}\simeq 0.53$ in the case of a RPP.}}
\label{fig:std}
\end{figure}
The level of clustering can be further quantified using the standard deviation of the normalized \vor{} areas $\sigma_\mathcal{V}=\sqrt{\langle \mathcal{V}^2 \rangle-1}$, plotted in Fig.~\ref{fig:std}.
For neutrally buoyant particles we find a constant value $\sigma_\mathcal{V} \simeq 0.53$, which is the expected value for a RPP. 
For heavy particles we find $\sigma_\mathcal{V}>0.53$ for all the experimental configurations investigated, revealing the presence of clustering.
This result is in agreement with previous measurements~\cite{guala:2008} that find clustering for large ($\phi\simeq4$) and heavy ($\Gamma=1.4$) particles.
We find the clustering level to globally decrease as $St_\text{h}$ and/or $R_\lambda$ increase, 
with no hint of maximum clustering for $St_\text{h}$ around unity.
This is contrary to common observations: 
for numerical Stokesian models~\cite{bec:2007,coleman:2009} a maximum is found for $St\simeq0.6$ ; for experiments with small particles~\cite{monchaux:2010,obligado:2011} a peak is observed for $St\gtrsim1$ while~\cite{fessler:1994} found a mild maximum for $St\simeq1$.
If a maximum of clustering exists in our case (which is reasonable assuming that tracer behavior is to be recovered for $St_\text{h}\rightarrow0$), the peak would be at $St_\text{h}<0.25$.
However, the limit $St_\text{h}\rightarrow0$ (\emph{i.e.} $R_\lambda\rightarrow0$) could not be explored here due to the settling effects at low $R_\lambda$.
The clustering properties (Stokes number dependence and clusters geometry) for such finite-size and heavy particles go beyond the scope of the present research and will be investigated in future experiments.

Two important conclusions can be drawn from these results. 
(i)~While inertial Stokesian models predict clustering within the explored range of St, this is not observed in the specific case of finite-size neutrally buoyant particles. Consequently, despite they do not behave as tracers, such particles are clearly not of the \emph{inertial} class. The absence of clustering also supports experimental results on the dynamics of finite-size neutrally buoyant particles~\cite{qureshi:2007,brown:2009,volk:2011} suggesting that simple time-response effects are not sufficient to describe the particle/flow interaction and that other mechanisms (such as the role of pressure increments at the scale of the particle) are to be accounted.
(ii)~Subsequently, the Stokes number by itself cannot be taken as sufficient to characterize clustering as we have shown that particles with similar Stokes number may or not exhibit preferential concentration.
Note that the limitation of the Stokes number has previously been shown for dynamical properties (rather than spatial distribution properties as illustrated here) of both isolated particles in turbulent flows~\cite{calza:2008,qureshi:2008} and two-way coupling effects~\cite{lucci:2011}. These observations combined to ours stimulate the need for further investigations on the possible connection between dynamical features and preferential sampling of particles (including for instance turbophoresis and stick-sweep mechanisms~\cite{goto:2008}, but also ergodic mechanisms~\cite{gustavsson:2011}), by coupling \vor{} analysis of particles distribution to Lagrangian tracking~\cite{monchaux:2010,tagawa:2012}.\\
We currently investigate which set of parameters $f(\phi,\Gamma, R_\lambda,...)$ drives the finite-size particle clustering, in particular by exploring how clustering is affected when Stokes number is varied (using particles with different sizes) at a fixed Reynolds number.

\begin{acknowledgments}
This work is partially supported by the ANR project TEC.
We thank D. Le Tourneau for the manufacturing of the LEM.
\end{acknowledgments}

\bibliography{biblio}

\end{document}